\documentstyle[12pt]{article}
\begin{document}
\tolerance=5000
\def\be{\begin{equation}}
\def\ee{\end{equation}}
\def\bea{\begin{eqnarray}}
\def\eea{\end{eqnarray}}
\def\nn{\nonumber \\}
\def\cF{{\cal F}}
\def\det{{\rm det\,}}
\def\Tr{{\rm Tr\,}}
\def\e{{\rm e}}
\def\etal{{\it et al.}}
\def\erp2{{\rm e}^{2\rho}}
\def\erm2{{\rm e}^{-2\rho}}
\def\er4{{\rm e}^{4\rho}}
\def\etal{{\it et al.}}
\def\gsim{\ ^>\llap{$_\sim$}\ }
\def\gd{g^\dagger}
\newcommand{\inv}[1]{\left[#1\right]_{\mbox{inv}}}
\newcommand\DS{D \hskip -3mm / \ }
\def\ds{\left( 1 + {M \over \lambda}
\e^{\lambda(\sigma^- - \sigma^+ )}
\right)}

\  \hfill
\begin{minipage}{3.5cm}
NDA-FP-55 \\
January 1999 \\
\end{minipage}

\

\vfill

\begin{center}

{\large\bf Conformal anomaly for 2d 
and 4d dilaton coupled spinors}

\vfill

{\large\sc Peter van Nieuwenhuizen}\footnote{
e-mail : vannieu@insti.physics.sunysb.edu}

{\large\sc Shin'ichi Nojiri$^{\clubsuit}$}\footnote{
e-mail : nojiri@cc.nda.ac.jp}
and
{\large\sc Sergei D. Odintsov$^{\spadesuit}$}\footnote{
e-mail :
odintsov@tspi.tomsk.su}

\vfill

{\large\sl Institute for Theoretical Physics \\
State University of New York \\
Stony Brook, NY 11794-3840, USA}

{\large\sl $\clubsuit$ Department of Mathematics and Physics \\
National Defence Academy \\
Hashirimizu Yokosuka 239, JAPAN}

{\large\sl $\spadesuit$
Tomsk Pedagogical University \\
634041 Tomsk, RUSSIA}

\vfill

{\bf ABSTRACT}

\end{center}

We study quantum dilaton coupled spinors in two
and four dimensions. Making classical transformation of metric, 
dilaton coupled spinor theory is transformed to minimal spinor theory 
with another metric and in case of 4d spinor also in the background 
of the non-trivial 
vector field. This gives the possibility to calculate 2d and 4d 
dilaton 
dependent conformal (or Weyl) anomaly in easy way. Anomaly induced
effective action for such spinors 
is derived. 
In case of 2d, the effective action reproduces, without any extra 
terms, the term added by hands 
in the quantum correction for RST model, which is exactly solvable.
For 4d spinor the chiral anomaly which depends explicitly from dilaton 
 is also found. As some 
application 
we discuss SUSY Black Holes in dilatonic supergravity with WZ type 
matter and Hawking radiation in the same theory. As another application 
we investigate spherically reduced Einstein gravity with 
 2d dilaton coupled fermion anomaly induced effective action and show the 
existence of quantum corrected Schwarszchild-de Sitter (SdS) (Nariai) BH 
with multiple horizon. 

\newpage

\section{Introduction}

Spherical reduction of Einstein gravity (see, for example \cite{8}) 
with minimal matter leads to some 2d dilatonic gravity (for most 
general action, see \cite{9}) with 2d dilaton coupled matter. 
Then, applying $s$-wave approximation one is forced to 
study 2d quantum dilaton coupled theory. Having in mind mainly 
cosmological applications, conformal anomaly may be one of the most 
important objects to study in such theory. For 2d dilaton coupled 
scalar the conformal (or Weyl) anomaly has been found in ref.\cite{1} 
and later in refs.\cite{2,3,NO2,4}. Using it, one can find 
anomaly induced, dilaton dependent effective action \cite{2,3,NO2} 
(see also later work \cite{7} where this effective action has been 
rederived). Anomaly induced effective action has been applied to 
investigate the following remarkable problems: Hawking radiation 
\cite{13} in dilatonic supergravity \cite{NO2},
anti-evaporation of multiple horizon black holes \cite{6,10},
discussion of semi-classical energy-momentum tensor in the 
presence of dilaton \cite{12}, etc. 

In the present paper we investigate the conformal anomaly 
for 2d and 4d dilaton 
coupled spinors. Naive transformation of quantum spinor 
(inclusion of dilatonic function to fermion) seems to indicate 
that no dilaton dependent terms appear in conformal anomaly for 
dilaton coupled spinor \cite{1,NO2,IO}. Unfortunately it is not quite
correct and Jacobian of quantum transformation is lost in this way. 
In this work, making only classical transformation of external 
classical gravitational field, we map dilaton coupled fermion to 
minimal fermion in the another classical gravitational background. 
(No transformation of quantum fields is made). Then, calculation of 
conformal anomaly may be done in quite straitforward way. 
In case of 2d, the anomaly induced effective action 
is derived. The effective action reproduces 
the term added by hands 
in the quantum correction for RST model\cite{RST}, 
which is exactly solvable. Adding such effective action to
reduced Einstein gravity, i.e. working in large $N$ and $s$-wave 
approximation we derive quantum corrected SdS (Nariai) BH.
The structure of theory under investigation shows the possibility 
of anti-evaporation of such BH in analogy with results of refs.
 \cite{6,10}.

The paper is organised as follows. In the next section we 
calculate the conformal anomaly and induced action for 2d dilaton 
coupled fermion. Comparizon with 2d scalar is made.
 The possibility of realization for quantum part 
of RST model as fermion anomaly induced effective action 
is shown. Quantum corrected SdS BH is also constructed.
 
Section 3 is devoted to short discussion of BHs in dilatonic 
supergravity with WZ type matter. Energy-momentum tensor 
components are found and Hawking radiation is derived. In section 4 
we evaluate conformal and chiral anomaly as well as induced action 
for 4d dilaton coupled fermion. Its explicit dilaton dependence 
is demonstrated. Finally, some discussion is 
presented in last section.

\section{Conformal anomaly for 2d dilaton coupled spinor}

Let us start from 2d dilaton coupled spinor Lagrangian:
\be
\label{i}
L=\sqrt{-g}f(\phi)\bar\psi \gamma^\mu \partial_\mu \psi
\ee
where $\psi$ is 2d Majorana spinor, $f(\phi)$ is an 
arbitrary function and $\phi$ is dilaton.

Let us make now the following classical transformation of background 
field $g_{\mu\nu}$:
\be
\label{ii}
g_{\mu\nu}\rightarrow f^{-2}(\phi)\tilde g_{\mu\nu}\ .
\ee
Then it is easy to see that $\gamma^\mu(x)\rightarrow f(\phi)
\tilde\gamma^\mu(x)$ and in terms of new classical metric we obtain usual, 
non-coupled with dilaton (minimal) Lagrangian for 2d spinor:
\be
\label{iii}
L=\sqrt{-\tilde g}\bar\psi\tilde\gamma^\mu\partial_\mu\psi\ .
\ee
The conformal anomaly for Lagrangian (\ref{iii}) is well-known:
\be
\label{iv}
\sqrt{-\tilde g}T={\sqrt{-\tilde g} \over 24\pi}\left\{{1 \over 2}
\tilde R\right\} \ .
\ee
Now, transforming Eq.(\ref{iv}) to original variables:
\be
\label{v}
\tilde g_{\mu\nu}=f^2(\phi)g_{\mu\nu}\ ,\ \ 
\tilde R=f^{-2}(\phi)\left(R - 2 \Delta\ln f\right)
\ee
we get the following conformal anomaly for dilaton coupled Majorana 
spinor (\ref{i}):
\bea
\label{vi}
\sqrt{-g}T &=& {\sqrt{-g} \over 24\pi}\left[{1 \over 2}R - 
\Delta \ln f \right] \nn
&=& {\sqrt{-g} \over 24\pi}\left[{1 \over 2}R - {f' \over f}
\Delta f - {\left(f''f - {f'}^2 \right) \over f^2}
g^{\mu\nu}\partial_\mu\phi\partial_\nu\phi \right]\ .
\eea
Hence, we found the conformal anomaly for 2d dilaton coupled Majorana 
spinor.

Above result corrects the error in refs.\cite{1,NO2,IO} 
where conclusion was made that for dilaton coupled spinor 
no new dilaton dependent terms appear in conformal anomaly 
if compare with minimal spinor. Note also that for 2d dilaton 
coupled Dirac spinor the conformal anomaly is twice of 
eq.(\ref{vi}). 

In the conformal anomaly (\ref{vi}) dilaton dependent 
terms appear in the form of total derivative. In principle, 
it means that this term is ambiguious by physical reasons. 
Indeed, in two dimensions the analogue of Einstein action 
looks like:
\be
\label{vii}
S={1 \over G}\int d^2x \sqrt{-g}Rf(\phi)\ .
\ee
Now, there exists the following relation
\be
\label{viii}
g_{\mu\nu}{\delta \over \delta g_{\mu\nu}}\int
d^2x \sqrt{-g}R f(\phi)=\Delta f(\phi)\ .
\ee
In other words, by finite renormalization of gravitational 
action (\ref{vii}), we can always change the coefficient 
of $\Delta f$ term in conformal anomaly. So in 2d gravity this term 
is only fixed by the physical renormalization condition. 
That is why total derivative term of conformal anomaly for dilaton 
coupled scalar is ambigious.

Now we discuss anomaly induced effective action for dilaton coupled 
spinor. The derivation goes in the same way as it was for dilaton coupled 
scalar \cite{2,3}. Making the conformal transformation of the metric 
$g_{\mu\nu}\rightarrow \e^{2\sigma}g_{\mu\nu}$ in the trace anomaly, 
and using relation
\be
\label{ix}
\sqrt{-g}T={\delta \over \delta\sigma}W[\sigma]
\ee
one can find the anomaly induced effective action $W$. 
In the covariant, non-local form it may be written as following:
\bea
\label{x}
W&=&-{1 \over 2}\int d^2x \sqrt{- g} \left\{ 
{1 \over 96\pi}R{1 \over \Delta}R \right.\nn
&& \left. + \left(F_2(\phi) 
- {\partial F_3(\phi) \over \partial \phi} \right)
(\nabla^\lambda \phi)(\nabla_\lambda \phi) {1 \over \Delta}R 
+ R \int F_3(\phi) d\phi\right\}
\eea
where
\be
\label{xb}
F_2(\phi)= - {f''f - {f'}^2 \over 24\pi f^2}\ ,\ \ 
F_3(\phi)= - {f' \over 24\pi f}\ .
\ee
Note that coefficient of second term is actually zero as is easy to 
check. Hence, 
we got anomaly induced effective action for dilaton coupled spinor. 

An interesting thing is that the effective action (\ref{x}) 
exactly reproduces the effective action of RST model \cite{RST}, 
which is exactly solvable. The RST model is given by adding the 
quantum correction 
\be
\label{RSTq}
W_{RST}=-{1 \over 2}\int d^2x \sqrt{- g} \left\{ 
{N \over 48\pi}R{1 \over \Delta}R 
+ {N \over 24\pi}\phi R\right\}
\ee
to the action of the CGHS model \cite{CGHS}
\be
\label{CGHS}
S_{CGHS}={1 \over 2\pi}\int d^2x \sqrt{-g}\e^{-2\phi}
\left(R + 4 \nabla_\mu\phi \nabla^\mu \phi + 4\lambda^2\right)
\ .
\ee
In \cite{RST}, the second term in $W_{RST}$ (\ref{RSTq}) is added 
by hands. In the work by Bousso-Hawking \cite{2}, 
it has been found that 
the there appears the third term in (\ref{x}), which corresponds to 
the second term in (\ref{RSTq}), from the conformal anomaly of 
the dilaton coupled scalar fields but there always appears the 
second term in 
(\ref{x}), which makes the model not to be exactly solvable, 
again. As found here, however, if we consider only dilaton coupled 
($2N$) spinor fields as a matter to CGHS model \cite{CGHS}, 
the model becomes exactly solvable even when we include quantum 
corrections and has the exact quantum solutions, 
in the conformal gauge :
\be
\label{riv}
g_{\pm\mp}=-{1 \over 2}\e^{2\rho}\ ,\ \
g_{\pm\pm}=0\ ,
\ee
including the following quantum black hole solution
\bea
\label{RSTs}
&& \Omega=\chi = - {\lambda^2 \over \sqrt\kappa}x^+x^-
- {\sqrt\kappa \over 4}\ln\left(\lambda^2 x^+ x^-\right)\ ,\nn
&& \Omega\equiv{\sqrt\kappa \over 2}\phi + 
{\e^{-2\phi} \over \sqrt\kappa} \ ,\ \ 
\chi\equiv\sqrt\kappa \rho -{\sqrt\kappa \over 2}\phi + 
{\e^{-2\phi} \over \sqrt\kappa} \ ,\ \ 
\kappa\equiv {N \over 12}\ .
\eea
Here we neglect the quantum correction form the dilaton and 
graviton. Similarly, changing classical part of the model by
the spherically reduced action of Einstein gravity one can find 
2d BHs in such model with quantum correction. However,
now such 2d BHs will get 4d interpretation.

As an example we consider  
 spherically reduced Einstein gravity with the quantum correction 
(\ref{RSTq}) from 
$2N$ ($N$ in 4d) spinors. The equations of motion are given by 
the variation over $g_{\pm\pm}$, $g_{+-}$ and $\phi$. They have the 
following form in the conformal gauge (\ref{riv}) :
\bea
\label{ri}
0&=&{\e^{-2\phi} \over 4G}\left(
2\partial_r \rho\partial_r\phi + \left(\partial_r\phi\right)^2
-\partial_r^2\phi\right) \nn
&& -{N \over 12}\left( \partial_r^2 \rho - (\partial_r\rho)^2 \right)
-{N \over 6}\left( 2 \partial_r \rho \partial_r \phi
- \partial_r^2 \phi \right) + N t_0 \\
\label{rii}
0&=&{\e^{-2\phi} \over 4G}\left(\partial_r^2 \phi
- 2 (\partial_r\phi)^2 - \Lambda \e^{2\rho} 
+ \e^{2\rho+2\phi}\right) \nn
&& +{N \over 12}\partial_r^2 \rho -{N \over 6}\partial_r^2\phi \\
\label{riii}
0&=& -{\e^{-2\phi} \over 4G}\left(-\partial_r^2\phi
+(\partial_r\phi)^2
+\partial_r^2 \rho+ \Lambda \e^{2\rho}\right) 
+ {N \over 6} \partial_r^2\rho \ .
\eea
Now we investigate if there exists a Nariai or SdS type solution \cite{Na} 
 for  
above equations. For this purpose, we assume $\phi$ is a constant:
$\phi=\phi_0$.
Then Eqs.(\ref{rii}) and (\ref{riii}) can be rewritten as follows:
\bea
\label{rvi}
0&=&{3 \over NG}\left( -\Lambda \e^{-2\phi_0} + 1 \right)\e^{2\rho}
+ \partial_r^2\rho \\
\label{rvii}
0&=&{3\Lambda \e^{-2\phi_0} \over 2NG}\left( 
{3 \e^{-2\phi_0} \over 2NG} - 1 \right)^{-1}\e^{2\rho}
+ \partial_r^2\rho \ .
\eea
Comparing (\ref{rvi}) with (\ref{rvii}), we obtain
\be
\label{rviii}
\e^{-2\phi_0}={NG \over 6}+{1 \over 2\Lambda}
\pm\sqrt{{1 \over 4\Lambda^2} - {NG \over 2\Lambda}
+ {N^2G^2 \over 36}}\ .
\ee
The sign $\pm$ in (\ref{rviii}) should be $+$ if we require the 
solution coincides with the classical one 
$\e^{-2\phi_0}={1 \over \Lambda}$ in the 
classical limit of $N\rightarrow 0$.
Then from (\ref{rvi}) or (\ref{rvii}), we find
\be
\label{rix}
\e^{2\rho}={2C \over R_0}{1 \over \cosh^2\left(r\sqrt{C}\right)}\ .
\ee
Here $C>0$ is a constant of the integration and $R_0$ is 2d scalar 
curvature, which is given by
\be
\label{rx}
R_0=-2\e^{-2\rho}\partial_r\rho
=\Lambda - {3 \over NG}\pm {6\Lambda \over NG}
\sqrt{{1 \over 4\Lambda^2} - {NG \over 2\Lambda}
+ {N^2G^2 \over 36}}\ .
\ee
(The sign $\pm$ corresponds to (\ref{rviii}).)
It is straightforward to check that the solution in (\ref{rviii}) and 
(\ref{rix}) satisfies Eq.(\ref{ri}).
The 4d curvature $R_4$ is also given by
\be
\label{rxi}
R_4=R_0 + 2\e^{2\phi_0}
= {3\Lambda \over 2} - {5 \over 2NG}\pm {3\Lambda \over NG}
\sqrt{{1 \over 4\Lambda^2} - {NG \over 2\Lambda}
+ {N^2G^2 \over 36}}\ .
\ee
 Above spherically reduced Einstein gravity with fermion 
quantum correction (in large $N$ and $s$-wave approximation)
 acquires the structure carefully studied  
in the model by Bousso-Hawking \cite{10}. It is easy to repeat literally  
the same investigation as in ref.\cite{10} and to show the possibility 
of anti-evaporation of Nariai BH due to quantum spinors.
(Note that ref.\cite{10} is dealing with only quantum dilaton coupled
scalars).
This study is even simpler as two terms of induced effective action for 
2d dilaton coupled scalars simply do not appear for the spinor case.
So it is not necessary to adopt approximation where these two terms disappear 
for the study of quantum equations of motion.

For comparison we give here also conformal anomaly and induced 
effective action for dilaton coupled scalar $a$ with Lagrangian:
\be
\label{xi}
L=f_s(\phi)g^{\mu\nu}\partial_\mu a\partial_\nu a \ .
\ee
Using results of refs.\cite{1,2}, one can write down:
\be
\label{xii}
\sqrt{-g}T={\sqrt{-g} \over 24\pi}\left\{R
-3\left({f''_s \over f_s }
- {{f'_s}^2 \over 2f_s^2}\right)(\nabla^\lambda\phi)(\nabla_\lambda\phi)
-3{f_s' \over f_s}\Delta\phi \right\}\ .
\ee
Anomaly induced effective action is given by Eq.(\ref{x}), where 
coefficient of the first term is ${1 \over 48\pi}$ (i.e. twice bigger), 
and
\be
\label{xiii}
F_2(\phi)= - {1 \over 8\pi}\left({f''_s \over f_s} 
- {{f'_s}^2 \over 2f_s^2}\right)\ , \ \ \ 
F_3(\phi)=- {1 \over 8\pi}{f'_s \over f_s}
\ee
Now, having at hands conformal anomaly for 2d dilaton coupled spinor 
and scalar we may discuss specific models.

\section{SUSY Black Holes and Hawking radiation}

We will start from the theory of the dilatonic supergravity with 
WZ type matter. The corresponding action has been constructed in 
ref.\cite{NO2} using modified form of tensor calculus (for the 
introduction, see \cite{N}).

Working on purely bosonic background (but still keeping fermions in 
matter sector) we may write the different versions of above dilatonic 
SG with matter. One of them is actually SUSY 
generalization \cite{NOi,PS} of CGHS model
\cite{CGHS} : 
\bea
\label{xiv}
L &=& - \left(\tilde V + {C(\phi) R \over 2} 
+ (Z+3\phi Z')(\nabla^\mu\phi)(\nabla_\mu\phi) \right. \nn
&& \left. - f(\phi)\sum_{i=1}^N 
\left[(\nabla_\mu a_i)(\nabla^\mu a_i) 
+ \bar\xi^i\gamma^\mu\partial_\mu\xi^i\right]\right)
\eea
where $\tilde V = -CS^2 - C'FS - ZF^2 - V'F$, 
$C(\phi)=2\e^{-2\phi}$, 
$Z(\phi)=4\e^{-2\phi}$, $V(\phi)=4\e^{-2\phi}$. 
Using the equations of motion with respect to the 
auxilliary fields, we obtain 
\be
\label{CGHSaf}
S=0\ ,\ \ F=-\lambda\ .
\ee 

We consider the situation 
where metric and dilaton are background fields 
and matter is quantized. Then, one can work in large $N$ 
approximation. The anomaly induced effective action due to $N$ scalars and 
spinors may be derived from Eqs.(\ref{x}), (\ref{xiii}) as the following:
\bea
\label{xv}
W&=&-{1 \over 2}\int d^2x \sqrt{- g} \left\{ 
{N \over 32\pi}R{1 \over \Delta}R \right. \nn
&& \left. - {N \over 16\pi}{{f'}^2 \over f^2} 
(\nabla^\lambda \phi)(\nabla_\lambda \phi) {1 \over \Delta}R 
- {N \over 6\pi}R\ln f\right\}\ .
\eea
Due to missing fermionic contribution the coefficient of last term 
in (\ref{xv}) is slightly different if compare with the result written 
in \cite{NO2}. 

The total effective action is given by sum of anomaly induced effective 
action $W$ and some conformally invariant functional which may be 
defined in Schwinger-De Witt expansion. The leading term of this expansion  
 for fermion is actually zero unlike the scalar. There are indications that
 this conformal invariant functional is exactly zero for dilaton coupled
spinor, 
 as it is impossible to construct conformal invariant combinations from
dilaton 
and flat derivatives (no massive coupling constants) with more than two
derivatives.
The corresponding SD coefficient on flat background for spinor is total
derivative 
 as is shown above. 
The exact solvability of RST model indicates the same.

We investigate the correction from 
the previous results in Ref.\cite{NO2} due to the missing 
fermion contribution. 
As a matter fields, we use dilaton coupled matter supermultiplet, 
which is natural as a toy model of the 4d models but is different 
from the original CGHS model.  
We are interested in the change of the estimation 
of the back-reaction from such a matter 
supermultiplet to black holes and Hawking radiation 
working in large-$N$ approximation.
Since we are interested in the vacuum (black hole) 
solution, we consider the background where 
matter fields, the Rarita-Schwinger field and 
dilatino vanish.

In the superconformal gauge, 
the equations of motion in \cite{NO2} obtained by the variation 
over $g^{\pm\pm}$, $g^{\pm\mp}$ and $\phi$ are slightly changed.
We should also note that there is, in general, a contribution from 
the auxilliary fields to $T_{\pm\mp}$ besides the 
contribution from the trace anomaly.

In large-$N$ limit, where classical part can be ignored, field equations 
become simpler and we can find the explicit solutions:
\be
\label{phi}
h(\phi)= \int d\rho {{4 \over 3}  \pm 
\sqrt{{16 \over 9} + \rho} \over \rho} \ ,\ \ \ 
\rho=-{16 \over 9} + 
\left(\rho^+(x^+) + \rho^-(x^-)\right)^{2 \over 3}\ .
\ee
Here $\rho^\pm$ is an arbitrary function of $x^\pm = t \pm x$. 
Comparing with the results in \cite{NO2}, the coefficients in 
(\ref{phi}) are slightly changed but the essential
behavior does not change. For example, the scalar curvature is given by
\bea
\label{sR}
R&=&8\e^{-2\rho}\partial_+\partial_-\rho \nn
&=& -{4\e^{-2\left\{
-{16 \over 9}+\left(\rho^+(x^+)+\rho^-(x^-)\right)^{2 \over 3}
\right\}} \over 
\left(\rho^+(x^+)+\rho^-(x^-)
\right)^{{4 \over 3}}}{\rho^+}'(x^+){\rho^-}'(x^-)\ .
\eea
Note that when $\rho^+(x^+)+\rho^-(x^-)=0$, there is a curvature
singularity. 
Especially if we choose
\be
\label{Krus}
\rho^+(x^+)={r_0 \over x^+}\ ,\ \ 
\rho^-(x^-)=-{x^- \over r_0}
\ee
there are curvature singularities at $x^+x^-=r_0^2$ and 
horizon at $x^+=0$ or $x^-=0$. 
Hence we got black hole solution in the model under discussion.
The asymptotically flat 
regions are given by $x^+\rightarrow +\infty$ ($x^-<0$) 
or $x^-\rightarrow -\infty$ ($x^+>0$).

We now consider Hawking radiation. 
We investigate the case that $f(\phi)=\e^{\alpha\phi}$
$(h(\phi)=\alpha\phi)$.
Hawking radiation can be obtained by 
substituting the classical black hole solution
which appeared in the original CGHS model \cite{CGHS} 
\be
\label{sws}
\rho=-{1 \over 2}\ln \left(1 + {M \over \lambda}
\e^{\lambda (\sigma^--\sigma^+ )} \right) \ , \ \ 
\phi=- {1 \over 2}\ln \left( {M \over \lambda}
+ \e^{\lambda(\sigma^+ - \sigma^-)} \right)  
\ee
(where $M$ is the mass of the black hole and we used asymptotic flat 
coordinates)
into the quantum part of the energy momentum tensor. We  
 use eq.(\ref{CGHSaf}). Then we find that when 
$\sigma^+\rightarrow +\infty$, the energy momentum tensor behaves as
\be
\label{asT}
T^q_{+-}\rightarrow 0 \ ,\ \ \ 
T^q_{\pm\pm}\rightarrow {N\lambda^2 \over 16}
\alpha^2 + t^\pm(\sigma^\pm)\ .
\ee
Here $t^\pm (\sigma^\pm)$ is a function which is determined by the 
boundary condition. 
In order to evaluate $t^\pm(\sigma^\pm)$, 
we impose a boundary condition that there is no incoming energy.
This condition requires that $T^q_{++}$ should vanish at the past 
null infinity   
($\sigma^-\rightarrow -\infty$) and if we assume $t^-(\sigma^-)$ is 
black hole mass independent, $T^q_{--}$ should also vanish at 
the past horizon ($\sigma^+\rightarrow -\infty$) after taking 
$M \rightarrow 0$ limit. Then we find
\be
\label{t-}
t^-(\sigma^-)=-{N\lambda^2\alpha^2 \over 16}
\ee
and one obtains
\be
\label{rad}
T^q_{--}\rightarrow 0
\ee
at the future null infinity ($\sigma^+\rightarrow +\infty$). 
Eqs.(\ref{asT}) and (\ref{rad}) might tell that there is no 
Hawking radiation in the dilatonic supergravity model 
under discussion when quantum back-reaction of 
matter supermultiplet in large-$N$ approach 
is taken into account as in \cite{NO2}. (That indicates that 
above black hole is extremal one).
 From another side since we work in strong coupling regime 
it could be that new methods to study Hawking radiation should be
developed.

\section{Conformal and chiral anomaly for 4d dilaton coupled spinor}

We consider now 4d dilaton coupled fermion which may appear 
as the result of spherical reduction of higher dimensional 
minimal spinor. The corresponding Lagrangian 
may be taken as follows:
\be
\label{Vi}
L=\sqrt{-g}f(\phi)\bar\psi \gamma^\mu(x) \nabla_\mu \psi
\ee
where $\nabla_\mu=\partial_\mu + {1 \over 2}\omega^{ab}_\mu\sigma_{ab}$ 
unlike the case of 2d Majorana
spinor. Note that the action (\ref{Vi}) is conformally invariant.

Let us make the following transformation of background gravitational field:
\be
\label{Vii}
g_{\mu\nu}=\e^{2\sigma(x)}\tilde g_{\mu\nu}\ ,\ \ 
\gamma^\mu(x)=\e^{-\sigma(x)}\tilde\gamma^\mu(x)\ ,\ \ 
\sqrt{-g}=\e^{4\sigma(x)}\sqrt{-\tilde g}\ .
\ee
It is easy to check then that 
\be
\label{Viii}
\gamma^\mu\nabla_\mu = \gamma^\mu\left(\partial_\mu 
+ {1 \over 2}\omega^{ab}_\mu\sigma_{ab}\right)
=\e^{-\sigma(x)}\tilde\gamma^\mu\left(\partial_\mu 
+ {1 \over 2}\tilde\omega^{ab}_\mu\sigma_{ab}
 + {3 \over 2}\partial_\mu\sigma(x)\right)\ .
\ee
(Note that review of conformal transformations in 4d gravity may be 
found in ref.\cite{fgn}.)
Let us select $\sigma(x)$ to satisfy
\be
\label{Viv}
\e^{3\sigma(x)}f(\phi)=1\ .
\ee
Then, Lagrangian (\ref{Vi}) after transformation (\ref{Vii}) 
takes the form:
\be
\label{Vv}
L=\sqrt{-\tilde g}\left[\bar\psi\tilde\gamma^\mu(x)\tilde\nabla_\mu\psi 
+ \bar\psi\tilde\gamma^\mu\tilde A_\mu\psi \right]
\ee
where we used $\partial_\mu\equiv \tilde\partial_\mu$ and 
$\tilde A_\mu = {3 \over 2}\partial_\mu\sigma(x)$. Note that 
 field strength for above vector field is equal to zero, that is 
 why no 
terms of the sort-square of field strength for above vector appear in
conformal anomaly. 
Hence, the calculation of $a_2$ Schwinger-De Witt coefficient in theory 
(\ref{Vi}) in curved spacetime with nontrivial dilaton is equivalent to 
the calculation of $a_2$ is an external gravitational field 
$\tilde g_{\mu\nu}$ (but no dilaton) and external vector field 
$\tilde A_\mu$.

Let us write the above Lagrangian as the following:
\be
\label{Vvi}
L=\sqrt{-\tilde g}\left[\bar\psi\tilde\gamma^\mu\tilde D_\mu \psi
\right]
\ee
where $\tilde D_\mu = \tilde \nabla_\mu + \tilde A_\mu$. We got the 
usual system: fermion in curved spacetime with the abelian external vector 
field. Conformal anomaly for such quantum (Dirac) fermion is 
well-known:
\be
\label{Vvii}
{\sqrt{-\tilde g} \over (4\pi)^2}\left\{{1 \over 20}\left( \tilde F 
+ {2 \over 3}\tilde\Box\tilde R \right) - {11 \over 360}\tilde G
\right\}
\ee
where
$\tilde F = \tilde R_{\mu\nu\alpha\beta}
\tilde R^{\mu\nu\alpha\beta}-2\tilde R_{\mu\nu}\tilde R^{\mu\nu}
+ {1 \over 3}\tilde R^2$, 
$\tilde G = \tilde R_{\mu\nu\alpha\beta}
\tilde R^{\mu\nu\alpha\beta}-4\tilde R_{\mu\nu}\tilde R^{\mu\nu}
+ \tilde R^2$. One can also present coefficients of conformal 
anomaly as 
 $b={1 \over 20(4\pi)^2}$ and $b'=-{11 \over 360(4\pi)^2}$.
And in principle, one can add $\tilde \Box \tilde R$ (total 
derivative) term with arbitrary coefficient to Eq.(\ref{Vvii}). 
It is known that coefficient of this term may be changed by finite 
renormalization of $R^2$-term in gravitational action, so it is 
ambigious.

Now, one can transform the relation (\ref{Vvii}) back to original 
metric tensor 
quantities:
\be
\label{Vviii}
\tilde g_{\mu\nu}\rightarrow \e^{-2\sigma}g_{\mu\nu}\ .
\ee
Then $\sqrt{-\tilde g}\tilde F \rightarrow \sqrt{-g} F$, 
$\tilde R\rightarrow \e^{2\sigma}[ R 
+ 6\Box \sigma - 6 (\nabla_\mu\sigma)(\nabla^\mu\sigma) ]$, 
$\sqrt{-\tilde g}\left(\tilde G - {2 \over 3}\tilde \Box \tilde R 
\right)\rightarrow \sqrt{-g}[ G - {2 \over 3} \Box R 
-4 \Box^2\sigma - 8R^{\mu\nu}\nabla_\mu\nabla_\nu\sigma 
+ {8 \over 3}R \Box \sigma - {4 \over 3}(\nabla^\mu R)
(\nabla_\mu\sigma)]$. 
Using the above relations, we may transform conformal anomaly to 
the original metric variables:
\bea
\label{Vix}
\sqrt{-g}T&=&{\sqrt{-g} \over (4\pi)^2}\left[{1 \over 20}F 
- {11 \over 360}\left( G - {2 \over 3}\Box R - 4 \Box^2\sigma \right. 
\right. \nn
&& \left. - 8R_{\mu\nu}\nabla_\mu\nabla_\nu\sigma + {8 \over 3}R\Box \sigma 
- {4 \over 3}(\nabla^\mu R)(\nabla_\mu \sigma)\right) \nn
&& +{2 \over 3}\left({1 \over 20} - {11 \over 360}\right)
\left\{ \Box R + 6\Box^2\sigma - 6 \Box \left((\nabla_\mu\sigma)
(\nabla^\mu\sigma) \right) \right.\nn
&& +2\Box\sigma R + 12\left(\Box\sigma\right)^2
-36\Box\sigma (\nabla_\mu\sigma)(\nabla^\mu\sigma) \nn
&& +2 \nabla_\mu \sigma \nabla^\mu R + 12 \nabla_\mu\sigma 
\nabla^\mu\Box\sigma - 12 \nabla_\mu\sigma \nabla^\mu 
\left((\nabla_\mu\sigma)(\nabla^\mu\sigma) \right) \nn 
&& \left.\left. -4 (\nabla_\mu\sigma)(\nabla^\mu\sigma) R 
+ 24 \left((\nabla_\mu\sigma)
(\nabla^\mu\sigma) \right)^2  \right\}\right]
\eea
where $\sigma = -{1 \over 3}\ln f$ depends explicitly from dilaton. 
Note that for Majorana spinor the conformal anomaly is given by 
${1 \over 2}$ of above expression.

Hence, we found conformal anomaly for 4d dilaton coupled spinor. 
It is easy to see that there are dilaton dependent 
contributions to conformal anomaly. One can also find 
anomaly induced effective action for dilaton coupled 4d spinor.
Starting from Eq.(\ref{Vvii}) for conformal anomaly in terms of 
tilded metric the corresponding anomaly induced action in 
this case is quite known (see \cite{RR}): 
\bea
\label{Vx}
W&=&-{1 \over 4b'}\int d^4x \sqrt{-\tilde g}
\int d^4x' \sqrt{-\tilde g'}\left[b\tilde F 
+ b'\left(\tilde G - {2 \over 3} \tilde \Box \tilde R\right) \right]_x \nn
&& \times \left[ 2\tilde \Box^2 + 4 \tilde R^{\mu\nu}
\tilde\nabla_\mu\tilde\nabla_\nu - {4 \over 3}\tilde R \tilde \Box 
+ {2 \over 3}\left(\tilde\nabla^\mu\tilde R\right)\tilde\nabla_\mu
\right]^{-1}_{xx'} \nn
&& \times \left[b\tilde F 
+ b'\left(\tilde G - {2 \over 3} \tilde \Box \tilde R\right) \right]_{x'} \nn
&& - {1 \over 18}(b+b')\int d^4x \sqrt{-\tilde g}\tilde R^2 \ .
\eea
It is trivial to substitute metric $\tilde g_{\mu\nu}=\e^{-2\sigma}
g_{\mu\nu}$, $\sigma=-{1 \over 3}\ln f(\phi)$ and rewrite above equation 
in terms of original metric and dilaton function. We do not do this in 
order to save the place.

As final remark, we note that similarly one can calculate the chiral 
anomaly for dilaton coupled spinor. Chiral anomaly for theory 
(\ref{Vv}) is known \cite{ef,N,et}:
\bea
\label{Vxi}
A_{1 \over 2}&=&{2i \over (4\pi)^2}\left[-{1 \over 48}
\epsilon^{\mu\nu\rho\sigma}{\tilde R^\xi}_{\zeta\mu\nu}
{\tilde R^\zeta}_{\xi\rho\sigma}\right] \nn
&=& {2i \over (4\pi)^2}\left[-{1 \over 48}
\epsilon^{\mu\nu\rho\sigma}\left\{{R^\xi}_{\zeta\mu\nu}
{R^\zeta}_{\xi\rho\sigma}
-16 {R^\xi}_{\rho\mu\nu}\nabla_\sigma\nabla_\xi\sigma \right. 
\right. \nn
&& \left.\left. -16 {R^\xi}_{\rho\mu\nu}\nabla_\sigma\sigma\nabla_\xi
\sigma 
-8 R_{\mu\nu\rho\sigma}\nabla_\alpha\nabla^\alpha \sigma
\right\}\right] 
\ .
\eea
Hence we found explicitly dilaton dependent corrections 
for chiral anomaly in the theory of 4d dilaton coupled spinor.

\section{Discussion}

In summary, we found the conformal anomaly and induced effective 
action for 2d and 4d dilaton coupled spinor as well as 4d chiral 
anomaly. The dilaton dependent part of conformal anomaly for 2d 
non-minimal spinor is shown to be the total derivative. As the 
result corresponding induced effective action has very simple form. 
It consists of two terms: Polyakov term and RST term which has been 
suggested as phenomenological, ad hoc term some time ago \cite{RST}. 
Hence, well-known RST model or, 
more exactly, some its simple modifications (in 
classical part) which were thought to be just 2d toy models acquire 
new remarkable interpretation. This model may be considered 
now as spherically reduced 4d gravity with spherically reduced 
minimal spinor ($s$-wave and large $N$ approximation is used). 
In other words, RST-like model may be applied to study 4d gravity (in 
above approximation). We give the example of
such sort showing the possibility to realize 
quantum corrected SdS (Nariai) BH in
reduced Einstein gravity 
with quantum correction due to large $N$ spinors.
Actually we work with 2d dilatonic BH which may be
re-interpreted as 4d BH without dilaton.
It is very interesting that this solution may be
also rewritten as cosmological solution which  
describes 4d Kantowski-Sachs Universe. Corresponding 
investigation will be presented elsewhere.

Similarly, one can consider anomaly induced effective action for 
4d dilaton coupled spinor to investigate quantum dilaton cosmology 
(no s-wave approximation then). In particulary, it would be very 
interesting to answer: can such fermions help in resolution of 
singularity problem via realization of non-singular Universes with 
non-trivial dilaton.

\ 

\noindent
{\bf Acknoweledgments.} We are grateful to M. Rocek and W. Siegel 
for helpful discussion. SDO would like to thank S.Hawking for 
stimulating discussion.


\begin{thebibliography}{99}
\bibitem{8} P. Hajicek, {\it Phys.Rev.} {\bf D30} (1984) 1178; \\
P. Thomi, B. Isaak and P. Hajicek, {\it Phys.Rev.} {\bf D30} 
(1984) 1168.
\bibitem{9}
S.D. Odintsov and I.L. Shapiro,
{\it Phys.Lett.} {\bf B263} (1991) 183; \\
T. Banks and M. O'Laughlin, {\it Nucl.Phys.} {\bf B362} (1991) 648.
\bibitem{1}
E. Elizalde, S. Naftulin and S.D. Odintsov,
{\it Phys.Rev.} {\bf D49} (1994) 2852.
\bibitem{2}
R. Bousso and S.W. Hawking, {\it Phys.Rev.} 
{\bf D56} (1997) 7788.
\bibitem{3}
S. Nojiri and S.D. Odintsov,
{\it Mod.Phys.Lett.} {\bf A12} (1997) 2083; 
{\it Phys.Rev.} {\bf D57} (1998) 2363;
\bibitem{NO2}
S. Nojiri and S.D. Odintsov, {\it Phys.Rev.} 
{\bf D57} (1998) 4847.
\bibitem{4}
T. Chiba and M. Siino, 
{\it Mod.Phys.Lett.} {\bf A12} (1997) 709; \\
S. Ichinose, {\it Phys.Rev.} {\bf D57} (1998) 6224; \\
W. Kummer, H.Liebl and D.V. Vassilevich,
{\it Mod.Phys.Lett.} {\bf A12} (1997) 2683; \\
J.S. Dowker,
{\it Class.Quant.Grav.} {\bf 15} (1998) 1881.
\bibitem{7}
W. Kummer and D.V. Vassilevich, preprint hep-th/9811092;
\bibitem{13}
S. Hawking,
{\it Comm.Math.Phys.} {\bf 43} (1975) 199;
\bibitem{6}
S. Nojiri and S.D. Odintsov, 
hep-th/9806055,
{\it Phys.Rev.} {\bf D} to appear;
{\it Mod.Phys.Lett.} {\bf A13} (1998) 2695;
\bibitem{10}
R. Bousso and S. Hawking,
{\it Phys.Rev.} {\bf D57} (1998) 2436;
\bibitem{12}
R. Balbinot and A. Fabbri, hep-th/9807123; \\
F.C. Lombardo, F.D. Mazzitelli and J.G.Russo, gr-qc/9808048;
\bibitem{IO} S. Ichinose and S.D. Odintsov, {\it Nucl.Phys.B},
to appear.
\bibitem{RST} J.G. Russo, L. Susskind and L. Thorlacius, 
{\sl Phys. Lett.} {\bf B292} (1992) 13.
\bibitem{CGHS} C.G. Callan, S.B. Giddings, J.A. Harvey 
and A. Strominger, {\it Phys. Rev.} {\bf D45} (1992) 
1005.
\bibitem{Na} H. Nariai, {\it Sci.Rep.Tohoku Univ.Ser.I} {\bf 35} 
(1951) 62.
\bibitem{N} P. van Nieuwenhuizen, {\it Phys.Rept.} 
{\bf 68} (1981) 189. 
\bibitem{NOi} Shin'ichi Nojiri and Ichiro Oda, 
{\it Mod.Phys.Lett.} {\bf A8} (1993) 53.
\bibitem{PS} Y. Park and A. Strominger, 
{\it Phys.Rev.} {\bf D47} (1993) 1566.
\bibitem{fgn} V. Faraoni, E. Gunzig and P. Nardone, 
gr-qc/9811047, 
{\it Fundamentals of Cosmic Physics}, to appear
\bibitem{RR} R.J. Reigert, {\it Phys.Lett.} {\bf B134} (1984)56; \\
E.S. Fradkin and A. Tseytlin, {\it Phys.Lett.} {\bf B134} (1984) 187; \\
I.L. Buchbinder, S.D. Odintsov and I.L. Shapiro, {\it Phys.Lett.} 
{\bf B162} (1985) 92; \\
I. Antoniadis and E. Mottola, {\it Phys.Rev.}  {\bf D45} 
(1992) 2013; \\
S.D. Odintsov, {\it Z.Phys.} {\bf C54} (1992) 531.
\bibitem{ef} T. Eguchi and P. Freund, {\it Phys.Rev.Lett.}
{\bf 37} (1976) 1251.
\bibitem{et} R. Endo and M. Takao, {\it Prog.Theor.Phys.}
{\bf 73} (1985) 803.
\end{thebibliography}
\end{document}